\documentclass{eptcs}
\providecommand{\event}{LAFM 2013} 
\usepackage{breakurl}             
\usepackage{graphicx}
\usepackage{url}

\def\halfthinspace{\relax\ifmmode\mskip.5\thinmuskip\relax\else\kern.8888em\fi}
\let \hts=\halfthinspace
\def \acompo{{\hts ; \hts}}

\def\({\left (}
\def\){\right )}
\def\<{\left <}
\def\>{\right >}
\def \[{\left[}
\def \]{\right]}

\title{The DynAlloy Visualizer}
\author{
Pablo Bendersky
\institute{Departamento de Computaci\'on, FCEyN, UBA\\Buenos Aires, Argentina}
\email{pbendersky@gmail.com}
\and
Juan Pablo Galeotti 
\institute{Saarland University \\ Saarbr\"ucken, Germany}
\email{galeotti@cs.uni-saarland.de}
\and
Diego Garbervetsky
\institute{Departamento de Computaci\'on, FCEyN, UBA\\Buenos Aires, Argentina}
\email{diegog@dc.uba.ar}
}
\def\titlerunning{The DynAlloy Visualizer (Tool Paper)}
\def\authorrunning{P. Bendersky, J. P. Galeotti \& D. Garbervetsky}
\begin{document}
\maketitle

\begin{abstract}
We present an extension to the DynAlloy tool to navigate DynAlloy counterexamples: the DynAlloy Visualizer.
The user interface mimics the functionality of a programming language debugger.
Without this tool, a DynAlloy user is forced to deal with the internals of the Alloy intermediate representation in order to debug a flaw in her model. 
\end{abstract}

\section{Introduction}

Alloy~\cite{Jackson:2006:SAL:1146359} is a formal specification language, which belongs to the class of the so-called model-oriented formal methods.
Among the key features that make Alloy an appealing modeling language to a wide community of users is its simple syntax, object oriented semantics and more importantly, its analyzability.
Alloy models can be automatically analyzed by simply declaring a bound on the number of elements for each domain.
We refer to this bound as the \emph{scope} of the analysis. 
Then, the Alloy model is translated into a propositional formula.
In turn, an off-the-shelf SAT-Solver is invoked to decide the validity of the given SAT-problem.
If the formula is satisfiable, then the semantic preserving translation ensures the original Alloy model has a counterexample.
Analogously, if the formula is unsatisfiable,  the translation ensures the Alloy model has no counterexamples, but \emph{within} the scope of the analysis previously selected by the user.
Both the Alloy tool and source code are publicly available for download\footnote{\url{http://alloy.mit.edu/alloy/download.html}}.

DynAlloy~\cite{DBLP:conf/icse/FriasGPA05} is an efficient extension of the Alloy language with procedural actions.
A DynAlloy user is able to declare atomic actions in terms of pre and postconditions using standard Alloy predicates.
Composite actions (namely programs) can be declared by combining these atomic actions and other simpler composite actions even further.
The syntax  for composite actions is given in Figure~\ref{grammaractions}.
A user can assert that if a given precondition holds, and a certain program execution ends, then the program execution led to a particular postcondition. 
In other words, since there is no guarantee of program termination, the semantics of DynAlloy assertions is that of partial correctness assertions.

\begin{figure}[h]
$$
\begin{array}{rclr}
program & ::= & \langle formula, formula \rangle(\overline{x}) & \mbox{``atomic action''}\\
       &  |  & formula ?                        & \mbox{``test''}\\
       &  |  & program + program                & \mbox{``non-deterministic choice''}\\
       &  |  & program \acompo program                & \mbox{``sequential composition''}\\
       &  |  & program^*                        & \mbox{``iteration''} \\
       &  |  & \langle program \rangle (\overline{x}) & \mbox{``invoke program''} \\
\end{array}
$$

\caption{Grammar for composite actions in
DynAlloy}\label{grammaractions}
\end{figure}

\begin{figure}
\begin{center}
\includegraphics[scale=0.80]{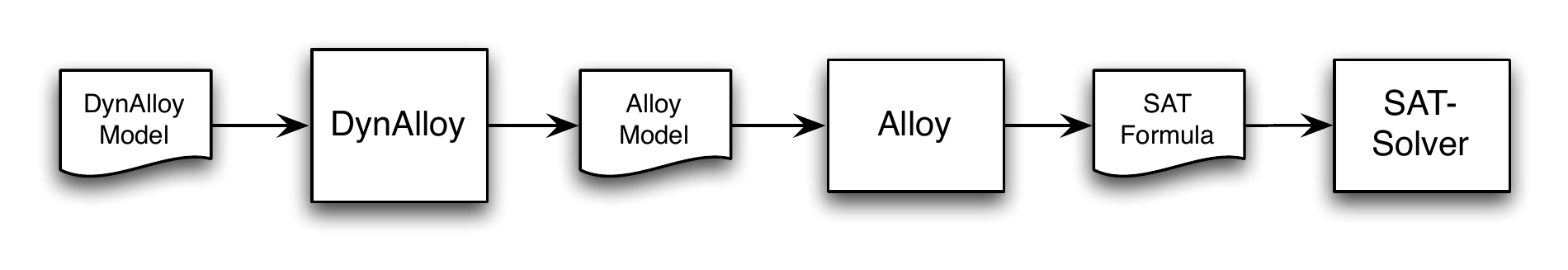}
\end{center}
\caption{A high-level view of the DynAlloy architecture}
\label{archDynAlloy}
\end{figure}

DynAlloy was devised with the explicit goal of allowing users to automatically analyze the written models.
In order to do so, the DynAlloy tool performs a semantic preserving transformation from a DynAlloy model into an Alloy model.
A high level description of DynAlloy's architecture is given in Figure~\ref{archDynAlloy}.
It is worth noticing that the \emph{iteration} composite action might lead to an infinite Alloy model for the general case.
To avoid this, DynAlloy restricts the analysis to traces of a fixed-length only.
As with the object domains, the user has to select this maximum fixed-length.
This is done by limiting the number of repetitions the iteration body might be executed.  

If we restrict ourselves to the architecture depicted in Figure~\ref{archDynAlloy}, a user obtains a rather binary result after analyzing a DynAlloy model.
Such answer might be expected when the given assertion is valid (always within the scope of analysis and the length of the traces under scrutiny).  
Nevertheless, if the assertion is invalid, the user is unable to understand in more depth how the intended model differs from the written one.
In other words, the user is unable to \emph{debug} the DynAlloy model.
 
In contrast, the Alloy tool includes a graphical user interface for displaying the counterexample found in terms of the relational semantics of the Alloy language.
This means that Alloy maps back the solution (namely the satisfying assignment) found by the SAT-solver from the propositional level, to the Alloy level.
Although exploring the Alloy counterexample appears as a more suitable solution than simply obtaining an error message, the user has to fully understand the internals of the translation of the model from DynAlloy to Alloy.

In this work we present an extension to the DynAlloy tool that allows users to navigate a counterexample in terms of the DynAlloy semantics rather than the Alloy semantics. 
The rest of the article is organized as follows.
In Section~\ref{sec-example} we present a motivational example. 
In Section~\ref{sec-dynalloy-gui} we describe our tool for visualizing DynAlloy counterexamples, as well as some implementation details.
Finally, in Section~\ref{conclusions} we conclude and discuss related work.

\section{A Motivational Example}\label{sec-example}

In order to illustrate the reader, let us a consider a simple example for specifying operations in singly linked lists. 
As we already stated, DynAlloy is an extension of the Alloy language. 
So, the user specifies different domains (namely, disjoint set of atoms) by declaring Alloy \emph{signatures}.
In the example we declare a singleton domain \emph{null} to represent an empty pointer.
Next, we declare two signatures, one for lists and the other for nodes contained in the lists:

\begin{verbatim}
one sig null {}
sig List {}
sig Node {}
\end{verbatim}

Each List element points to a Node element (or the null value) by means of field \emph{header}.
Similarly, a Node element points to the next Node element (or the null value) through field \emph{next}.
The count of elements within the list is stored using field \emph{size}. 
In order to allow updating the value of fields, \emph{head}, \emph{size} and \emph{next} fields are declared as functional variables instead of regular Alloy fields (whose semantics is static).
In the program body, the $a\rightarrow d$ denotes the ordered pair $\langle a, d \rangle$, and $++$ denotes the relational override between a functional relation and an ordered pair. 
The effect of such operator is to replace the previous value stored for $a$ with $d$ in the new value of the functional relation. 

\begin{small}  
\begin{verbatim}
program removeLast[thiz: List, header: List -> one(Node+null),
                   next: Node -> one(Node+null), size: List -> one Int]
var [curr: Node+null, prev: Node+null, newSize: Int]  {
  prev := null;
  curr := thiz.header;
  while isNotNull[curr] do {
    prev := curr; curr := curr.next
  };
  if isNotNull[prev] {
    header := header ++ (thiz->null);
    newSize := sub[thiz.size,1];
    size := size ++ (thiz->newSize)
  } else { skip }
}
\end{verbatim}
\end{small}  

As the reader may have noticed, \texttt{if} and \texttt{while} constructs are not valid productions in the grammar introduced in Figure~\ref{grammaractions}.
It is worth noticing that these more complex programming structures can be described using these basic logical constructs. 
More specifically, \texttt{if B \{P\} else \{Q\}} can be rewritten as the following DynAlloy program $B?\acompo P +(\neg B)?\acompo Q$. 
Similarly, \texttt{while B do \{P\}} can be expressed as $(B?; P)*\acompo(:B)?$.

Finally, the user declares a property of interest and the corresponding DynAlloy assertion.
In this case, she wants to state that the value stored in the \emph{size} field of the list agrees with the actual number of Node elements that are reachable from the \emph{header} field traversing the \emph{next} field. 
Such property can be compactly expressed by means of the Alloy reflexive transitive closure operator:
\begin{center}
\texttt{thiz.size=\#(thiz.header.*next-null)}
\end{center}

\begin{figure}
\begin{center}
\includegraphics[scale=0.40]{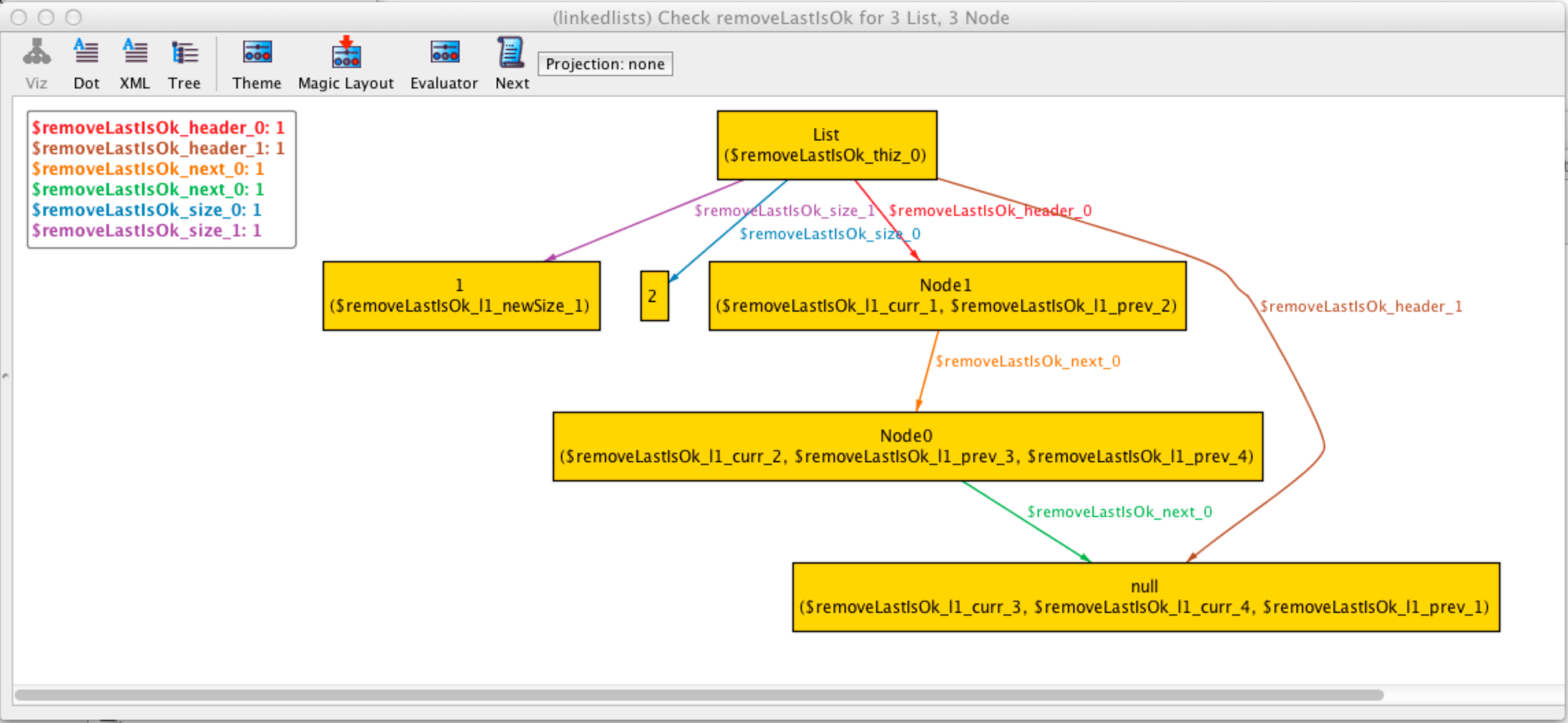}
\end{center}
\caption{The Alloy counterexample displayed using the standard visualization}
\label{alloygui1}
\end{figure}


For checking the correctness of the DynAlloy assertion, let us assume the user chooses the default scope of analysis (each signature can contain at most $3$ atoms and iterations can be exercised at most $3$ times).
Not very surprisingly, the analysis reports the assertion does not hold. 
In this scenario the only source for understanding the problem (apart from the DynAlloy model itself) is the Alloy counterexample reported by the Alloy tool (Figure~\ref{alloygui1}).
As the reader may notice, it is rather cumbersome to understand exactly what part of the DynAlloy model is flawed.
This is due to the fact that the counterexample is shown in terms of the \emph{internal} representation of the DynAlloy model into an Alloy model that the DynAlloy tool applies. 

\section{Debugging a DynAlloy model}\label{sec-dynalloy-gui}

In this work we present an extension to the DynAlloy tool that allows users to navigate a DynAlloy counterexample, the DynAlloy Visualizer.
The visualizer extension performs the following tasks:
\begin{itemize}
\item DynAlloy models are translated to Alloy models as presented in \cite{DBLP:conf/icse/FriasGPA05}, but preserving a mapping from DynAlloy to Alloy.
\item If a counterexample is found, the counterexample is lifted from the Alloy domain back into the DynAlloy realm using the collected mapping.  
\end{itemize}
 
\begin{figure}
\begin{center}
\includegraphics[scale=0.35]{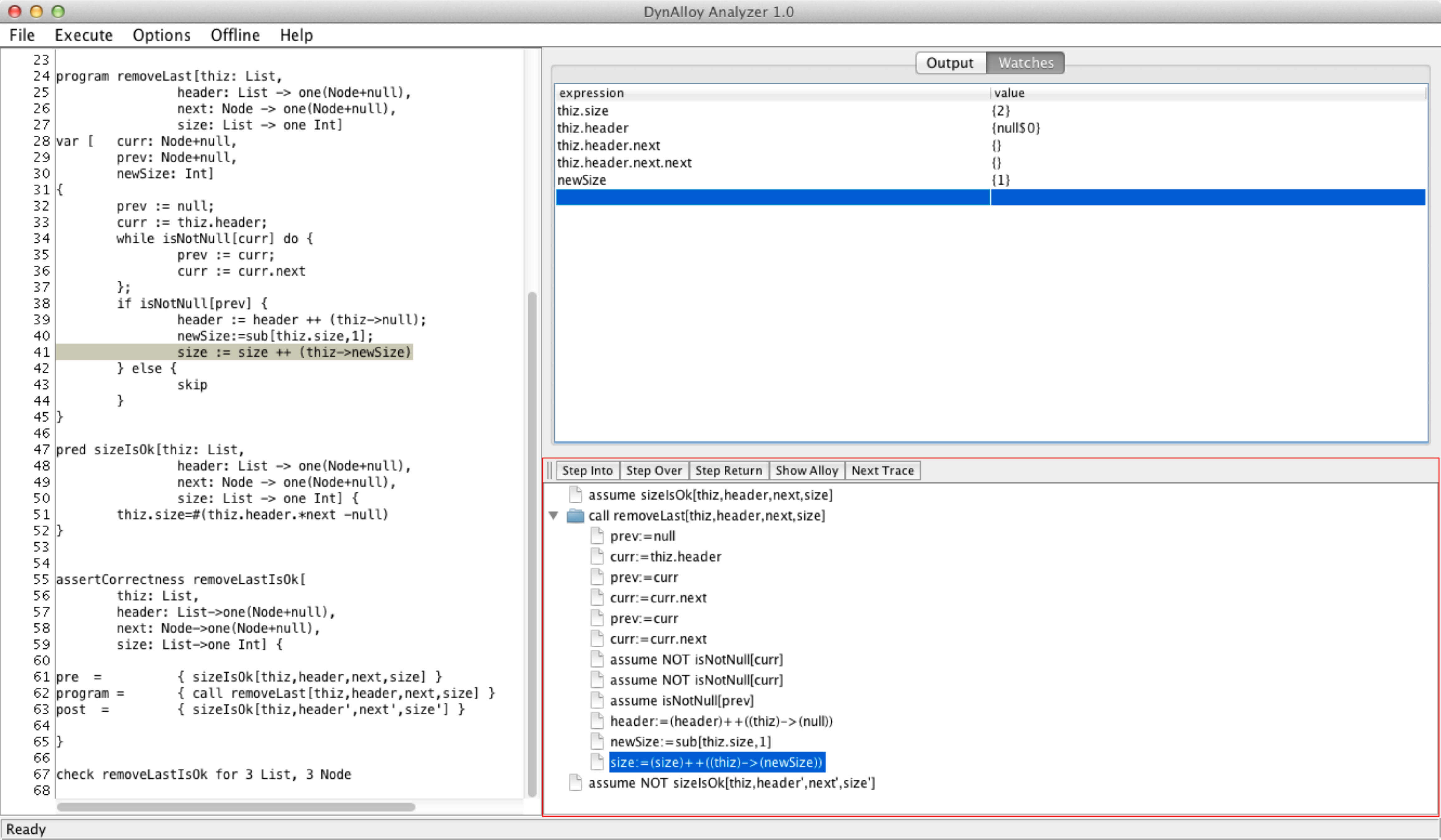}
\end{center}
\caption{The graphical-user interface of the DynAlloy Visualizer. 
The left pane allows the user to edit changes on the DynAlloy model. 
The right panes display the counterexample (if found).
The lower right pane shows the trace while the upper right pane displays the values of the selected expressions at the current step of the trace.}
\label{dynalloygui1}
\end{figure}

\begin{figure}
\begin{center}
\includegraphics[scale=0.60]{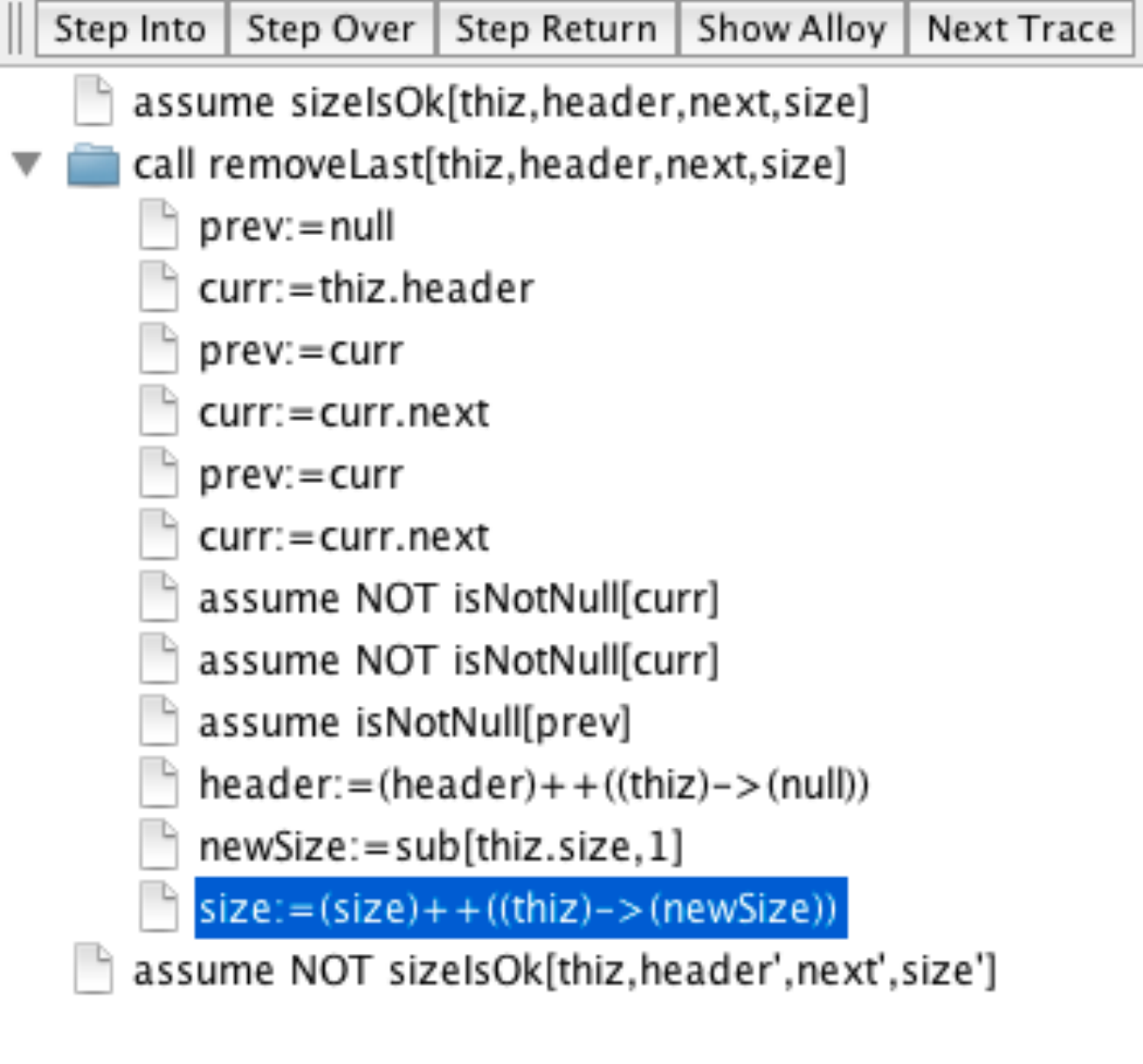}
\end{center}
\caption{The DynAlloy trace in detail. 
In the tree form view each node represents an action taken and parent nodes represent DynAlloy program calls.
The DynAlloy user is able to go backward and forward in the trace by simply clicking the new step.
}
\label{dynalloytrace1}
\end{figure}

Figure~\ref{dynalloygui1} shows the graphical-user interface of the DynAlloy Visualizer.
This interface allows the user to navigate the trace and inspect user-defined expressions and field values. 
Users can edit the DynAlloy model using the left pane.
Once editing is finished, she can launch a new analysis by hitting the \emph{Execute} menu option. 
In case a counterexample is found, the trace is presented on the lower right pane (Figure~\ref{dynalloytrace1}).
The counterexample trace is presented in a tree form where each node represents an action taken (an atomic action, an assignment, an \emph{if} decision, etc.) and parent nodes represent DynAlloy program calls.

On the upper right pane the values of selected variables, fields and user-defined Alloy expressions are displayed.
If the user navigates the counterexample trace (by going back and forth in the steps of the trace) these values are automatically updated to display their values at the current step.

In the example, the user can notice that although the size of the list correctly decreases from $3$ to $2$, the resulting list has no elements since the \emph{header} field points to null. 
In other words, instead of setting the \emph{header} field to null, what the program has to do at this point is set the value of \texttt{prev.next} to null. 
Once this change to the model is done, the new analysis returns no counterexample.

\subsection*{Implementation Details}

As already mentioned, while translating a DynAlloy model to the Alloy language, our tool builds a mapping from the source DynAlloy programs into the target Alloy formulas.
By using a bidirectional mapping we are also able to efficiently find the DynAlloy program for a given Alloy formula.
Luckily, the Alloy tool provides a programmatic interface for inspecting a given counterexample.
Each counterexample found leads to a fresh \texttt{A4Solution} instance with all the necessary information.
Following the translation presented in \cite{DBLP:conf/icse/FriasGPA05}, a disjunction in the target Alloy formula is introduced only in the case of the non-deterministic choice $+$.
Given the abstract syntax tree of the resulting Alloy formula, our tool evaluates each sub formula.
If the sub formula evaluates to true, then the original DynAlloy program is obtained, and the procedure is recursively applied on each component of the sub formula.
We implemented this algorithm generically by instantiating the \emph{Visitor} design pattern. 

A useful feature of Alloy Analyzer is the unsat core highlighting. 
This feature allows a user to see a (possibly minimal) subset of the model constraints from which the assertion follows.
Although the current version of our prototype does not offer such highlighting, the current DynAlloy architecture is suitable for supporting this feature in a future release.

\vspace{-0.5cm}
\section{Conclusions and Related Work}\label{conclusions}

In this work we have presented an extension to the DynAlloy tool for visualizing DynAlloy counterexamples.
We have also shown through a motivational example how this extension might help a user to debug her model without dealing with any intermediate representation. 

Among many other reasons, the success of Alloy as a lightweight formal method was based in helping its users to understand the counterexample provided by its back-end (in this case, an off-the-shelf SAT-Solver).
Many other tools follow this line.
The Boogie Verification Debugger~\cite{DBLP:conf/sefm/GouesLM11} provides plugins for visualizing counterexamples for VCC and Dafny users. 
JForge~\cite{Dennis09arelational} is a tool for Java bounded verification based on the Alloy language. 
The JForge plug-in allows the user to visualize the offending program trace, but described in an intermediate representation language.
TacoPlug~\cite{6229808} is an eclipse plugin for the bounded verifier TACO~\cite{DBLP:conf/issta/GaleottiRPF10}.
TacoPlug provides many debugging features such as heap memory graph visualization and Java trace navigation.
Since TACO uses DynAlloy as a backend, TacoPlug is a client of the Visualizer's API.

Our tool is implemented as an extension of the DynAlloy tool, which has been released as open source to the community. 
For more information on DynAlloy, please visit the website: \url{http://www.dc.uba.ar/dynalloy}
\vspace{-0.5cm}
\section*{Acknowledgments}
This work has been partially funded by CONICET, UBACyT-20020110200075/20020100100813, MinCyT PICT-2010-235/2011-1774/2012-0724, CONICET-PIP 11220110100596CO, MINCYT-BMWF AU/10/19, INRIA Associated Team ANCOME, and LIA INFINIS and MEALS 295261.
\vspace{-0.5cm}

\nocite{*}
\bibliographystyle{eptcs}
\bibliography{lafm2013}
\end{document}